\documentclass[aps,prl,preprint,amsfonts,nobibnotes,superscriptaddress,showpacs]{revtex4}

\usepackage{dcolumn}
\usepackage{amsmath}
\usepackage{amssymb}
\usepackage{graphicx}
\usepackage{bm}
\usepackage{upgreek,xspace}
\usepackage{color}
\usepackage{titlesec}
\DeclareMathOperator{\sign}{sign}

\begin{document}

\title{Exciton Control in a Room-Temperature Bulk Semiconductor with Coherent Strain Pulses}

\vspace{2cm}

\author{Edoardo Baldini}
	\affiliation{Department of Physics, Massachusetts Institute of Technology, Cambridge, Massachusetts, 02139, USA}

\author{Adriel Dominguez}
	\affiliation{Departamento Fisica de Materiales, Universidad del Pa\'is Vasco, Av. Tolosa 72, E-20018, San Sebastian, Spain}
	
\author{Tania Palmieri}
\affiliation{Laboratory of Ultrafast Spectroscopy, ISIC and Lausanne Centre for Ultrafast Science (LACUS), \'Ecole Polytechnique F\'ed\'erale de Lausanne (EPFL), CH-1015 Lausanne, Switzerland}
	
\author{Oliviero Cannelli}
\affiliation{Laboratory of Ultrafast Spectroscopy, ISIC and Lausanne Centre for Ultrafast Science (LACUS), \'Ecole Polytechnique F\'ed\'erale de Lausanne (EPFL), CH-1015 Lausanne, Switzerland}
	
\author{Angel Rubio}
	\affiliation{Departamento Fisica de Materiales, Universidad del Pa\'is Vasco, Av. Tolosa 72, E-20018, San Sebastian, Spain}
	\affiliation{Max Planck Institute for the Structure and Dynamics of Matter, Hamburg, Germany}
	
\author{Pascal Ruello}
	\affiliation{Institut des Mol\'ecules et Mat\'eriaux du Mans, UMR CNRS 6283, Le Mans Universit\'e, 72085 Le Mans, France}

\author{Majed Chergui}
	\affiliation{Laboratory of Ultrafast Spectroscopy, ISIC and Lausanne Centre for Ultrafast Science (LACUS), \'Ecole Polytechnique F\'ed\'erale de Lausanne (EPFL), CH-1015 Lausanne, Switzerland}

\date{\today}

\begin{abstract}
The coherent manipulation of excitons in bulk semiconductors via the lattice degrees of freedom is key to the development of acousto-optic and acousto-excitonic devices. Wide-bandgap transition metal oxides exhibit strongly bound excitons that are interesting for applications in the deep-ultraviolet, but their properties have remained elusive due to the lack of efficient generation and detection schemes in this spectral range. Here, we perform ultrafast broadband deep-ultraviolet spectroscopy on anatase TiO$_2$ single crystals at room temperature, and reveal a dramatic modulation of the exciton peak amplitude due to coherent acoustic phonons. This modulation is comparable to those of nanostructures where exciton-phonon coupling is enhanced by quantum confinement, and is accompanied by a giant exciton shift of 30-50 meV. We model these results by many-body perturbation theory and show that the deformation potential coupling within the nonlinear regime is the main mechanism for the generation and detection of the coherent acoustic phonons. Our findings pave the way to the design of exciton control schemes in the deep-ultraviolet with propagating strain pulses.
\end{abstract}


\maketitle

New perspectives in the field of excitonics have recently developed from the discovery of strongly bound excitons that persist at room temperature (RT) in several semicondcutors, including organics \cite{brutting2006introduction}, transition metal dichalcogenides \cite{schaibley2016valleytronics} and transition metal oxides \cite{Klingshirn, gogoi2015anomalous, ref:baldini_TiO2}. Despite their different origin, excitons in these classes of materials are strongly coupled to the lattice degrees of freedom. Indeed, since excitons can be viewed as quanta of electronic excitation energy travelling in the periodic crystal lattice, their motion is influenced by the fluctuating potential field due to lattice vibrations. On the fundamental aspect, exciton-phonon coupling is an intriguing type of boson-boson interaction that results in phenomena such as exciton self-trapping, spectral-weight transfers to phonon sidebands and Stokes-shifted emissions~\cite{ref:toyozawa}. On the practical side, identifying the specific modes (optical or acoustic) that couple strongly to the excitons paves the way to the control of the exciton properties through the tailored application of strain, pressure or photoexcitation.

Experimentally, the microscopic details of the exciton in the phonon field can be addressed via absorption and photoluminescence spectroscopy, since the shape and width of the optical spectra directly reflect the scattering of the exciton by lattice vibrations \cite{song2013self}. However, the information offered by these methods is mediated over all the coupled phonon modes. This calls for more advanced techniques that can yield information on the exciton-phonon coupling for specific lattice modes of interest in order to allow, in return, the phonon-selective control of the exciton properties. A powerful tool relies on setting a particular phonon mode out of equilibrium and monitoring the impact of the ionic motion on the exciton spectral features~\cite{ref:merlin, ref:ruello, mann2015probing}. This is possible by time- and energy-resolved optical spectroscopy, in which a system is first excited by an ultrashort laser pulse and the changes in the optical properties are tracked with a delayed optical probe covering the exciton lineshape \cite{yamamoto1994coherent, cerullo1999size, tyagi2010controlling, dworak2011coherent, ge2014coherent, bakulin2016real}. This approach also opens the door to the high-speed control of the exciton properties via ultrafast light excitation. So far, the coherent manipulation of excitons through the photoinduced ionic motion has led to exciton shifts as large as 10 meV in semiconductor nanostructures, and only at very low temperatures \cite{akimov2006ultrafast, scherbakov2007chirping}. At room temperature (RT), shifts of $<$ 1 meV have only been reached in quantum dots \cite{sagar2008size}, where the exciton-phonon coupling is enhanced by the low-dimensionality. This has posed serious limitations to the use of this approach for the design of efficient acousto-excitonic devices.

An alternative strategy involves the use of bulk semiconductors that are known to host strongly bound excitonic resonances at RT and simultaneously show strong electron-phonon coupling phenomena. An ideal candidate is the anatase polymorph of TiO$_2$, which is a superior system for several applications, ranging from photocatalysis and transparent conductive substrates to photovoltaic and sensors \cite{fujishima1972electrochemical, zhang2007niobium, bai2014titanium}. Its optical absorption spectrum is dominated by strongly bound excitons with an intermediate character between the Wannier-Mott and Frenkel regimes \cite{ref:baldini_TiO2}. These excitons are very robust against external perturbations, persisting at RT and high photodoped carrier densities, as well as in defect-rich nanoparticles and mesoporous films \cite{ref:baldini_TiO2, baldini2017interfacial, baldini2017exciton}. There exists a moderately strong coupling between the electronic degrees of freedom and polar longitudinal optical phonons via the Fr\"ohlich interaction, which gives rise to a significant polaronic dressing of the carriers at the bottom of the conduction band \cite{ref:moser, setvin2014direct}. Yet, the direct coupling between the excitonic states and acoustic modes is still unexplored. Here, using ultrafast broadband spectroscopy in the deep-UV, we reveal the signature of coherent acoustic phonons (CAPs) that couple directly to the c-axis exciton peak of anatase TiO$_2$. Thanks to the high photogenerated carrier densities that can be supported by this material, the amplitude of the reflectivity modulation produced by CAPs is among the largest ever reported, and indicates efficient generation/detection of acoustic modes. Furthermore, the strong exciton-phonon coupling produces a giant shift of the exciton peak, as large as 30-50 meV. We rationalize our findings within the framework of many-body perturbation theory, providing a complete quantitative treatment of such a strong exciton-phonon coupling in the nonlinear regime. It appears that the origin of the strong electron-phonon coupling is a consequence of the large deformation potential (DP) parameter, being itself determined by the intrinsic electronic properties of the material. Our results open perspectives for acousto-optic and acousto-excitonic RT applications of this widely known semiconductor.

High-quality anatase TiO$_2$ single crystals were grown by a chemical vapor-transport method and oriented via Laue diffraction to expose a (010) surface to the incoming radiation. Ultrafast broadband deep-UV spectroscopy was performed using the set-up described in Ref. \cite{ref:baldini_TiO2}. Many-body perturbation theory at the GW level and the Bethe-Salpeter Equation (BSE) \cite{ref:hedin1, ref:onida} was employed to compute the band structure and the dielectric response of the material. More details are provided in the Methods section.

\begin{figure}[t]
	\begin{center}
		\includegraphics[width=0.8\columnwidth]{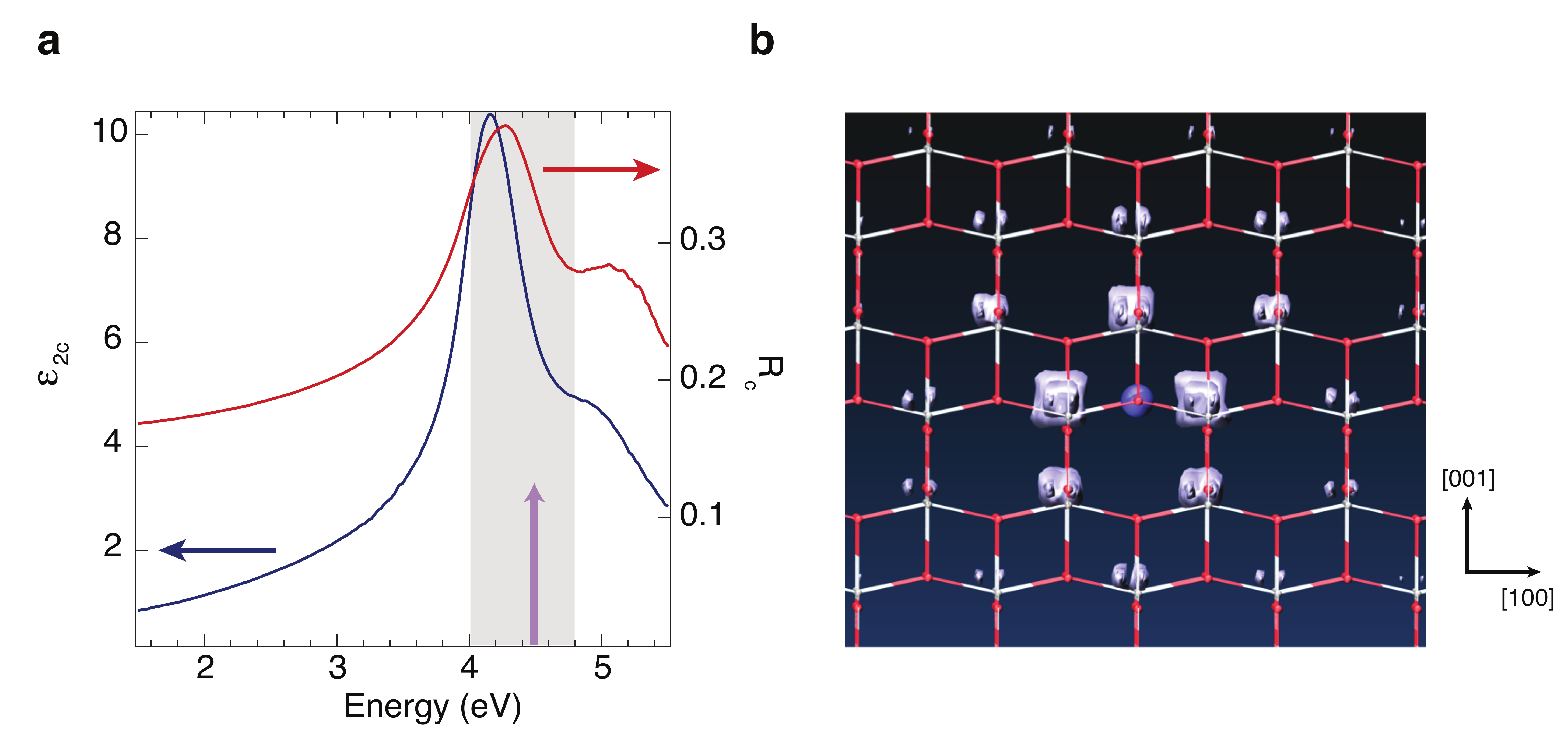}
		\caption{(a) Imaginary part of the dielectric function (blue curve) and reflectivity (red curve) of anatase TiO$_2$ measured at RT with the electric field polarized along the c-axis. The experimental data are obtained from Ref. \cite{ref:baldini_TiO2}, as measured by spectroscopic ellipsometry. The pump photon energy of 4.50 eV used for the pump-probe experiment is indicated by the violet arrow and the probed region is highlighted as a grey shaded area. (b) Wavefunction of the c-axis exciton of anatase TiO$_2$. The isosurface representation shows the electronic configuration when the hole of the considered excitonic pair is localized close to one oxygen atom. The coloured region represents the excitonic squared modulus wavefunction.}
		\label{fig:Fig1}
	\end{center}
\end{figure}

Figure 1(a) shows the imaginary part ($\epsilon_{2c}$) of the dielectric function at RT with light polarized parallel to the c-axis (blue trace), along with the material reflectivity (R) as derived from the dielectric function (red trace). These spectra are obtained from our spectroscopic ellipsometry data of Ref. \cite{ref:baldini_TiO2}. The $\epsilon_{2c}$ trace features a sharp peak at 4.15 eV, which is due to an excitonic transition \cite{ref:baldini_TiO2}. The binding energy of this collective excitation is $\sim$ 150 meV and its wavefunction (Fig. 1(b)) is fairly localized in all three directions, with an average Bohr radius of 0.7 - 2 nm. A weaker charge excitation lies around 5.00 eV and is ascribed to a resonant interband transition within the continuum. Consistently, R presents similar features at 4.26 eV and 5.15 eV, respectively. 

In our experiments, we excite the anatase TiO$_2$ crystal with an ultrashort laser pulse polarized along the c-axis. Its photon energy of 4.50 eV (violet arrow in Fig. 1(a)) lies above the exciton peak, in order to non-resonantly generate uncorrelated electron-hole pairs in the solid. The photoexcited carrier density is set to $N\sim$ 3.5 $\times$ 10$^{20}$ cm$^{-3}$. Subsequently, we monitor the relative changes in the material c-axis reflectivity ($\Delta$R/R) over a broad spectral range covering the exciton feature (grey shaded area in Fig. 1(a)). The time resolution of the experiments is $\sim$ 700 fs.

\begin{figure}[t]
	\begin{center}
		\includegraphics[width=0.8\columnwidth]{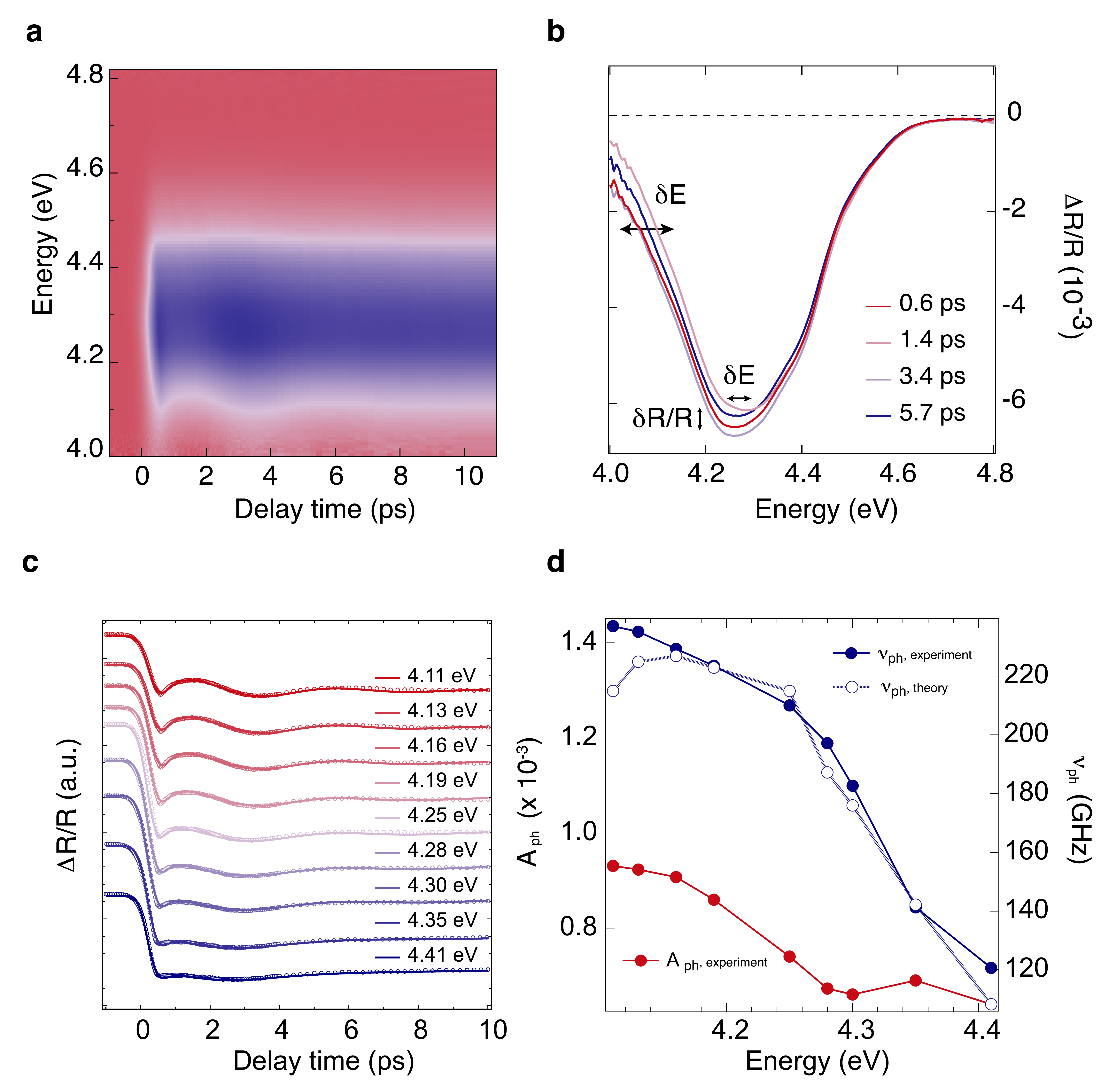}
		\caption{(a) Color-coded map of $\Delta$R/R at RT as a function of probe photon energy and time delay between pump and probe. Both pump and probe beams are polarized along the material c-axis. The pump photon energy is 4.50 eV. (b) Transient spectra of $\Delta$R/R for different time delays during the first 10 ps of the response. (c) Temporal traces of $\Delta$R/R for different probe photon energies (dotted lines), indicated in the label. The solid lines are fits to the experimental data. (d) Probe photon energy dependence of the amplitude ($\mathrm{A_{ph}}$) and frequency of the coherent oscillations ($\mathrm{\nu_{ph}}$) with a comparison between experiment and theory.}
		\label{fig:Fig2}
	\end{center}
\end{figure}

Figure 2(a) displays the color-coded map of $\Delta$R/R as a function of the probe photon energy and time delay between pump and probe. We observe a long-lived negative $\Delta$R/R response, which is due to the combination of phase-space filling and Coulomb screening (\textit{i.e.} exciton bleaching in the corresponding transient absorption signal) induced by the high photoexcited carrier densities \cite{ref:baldini_TiO2}. Importantly, this feature experiences a pronounced sinusoidal modulation during the first 10 ps, which is particularly evident around 4.10 eV. The $\Delta$R/R spectra at representative time delays are shown in Fig. 2(b). A closer inspection during the first 6 ps reveals that the exciton feature undergoes a dramatic modulation of its energy ($\delta$E as large as 30 meV at the exciton peak and 50 meV at the exciton low-energy tail) and intensity ($\delta$R/R $\sim$ 5\% of the total signal at the exciton peak). The time traces are displayed in Fig. 2(c) and show the large-amplitude coherent oscillation on top of a flat incoherent background. The oscillation frequency is much lower than those of coherent optical phonons \cite{portuondo2008ultrafast, ref:ishioka} (that cannot be accessed by the present time resolution), and depends on the probe photon energy. It is also strongly damped, vanishing after two periods. A global fit of the $\Delta$R/R temporal traces with a multiexponential function, convoluted with our instrument response function, and a damped sinusoidal term yields the evolution of the significant parameters of the response as a function of probe photon energy. The fitted curves are shown in Fig. 2(c) as solid lines on top of the experimental traces, demonstrating the accuracy of the global fit. The dependence of the oscillation amplitude (red trace) and frequency (blue trace) on the probe photon energy is plotted in Fig. 2(d). We observe that the amplitude peaks around 4.15 eV, \textit{i.e.} at the low-energy tail of the exciton resonance in the reflectance spectrum. The frequency also shows a complex dependence with the probe photon energy, which is not expected in the case of coherent optical phonons. This observable supports that the oscillations are due to longitudinal CAPs propagating along the [010] axis of the anatase TiO$_2$ single crystal, as demonstrated later. Remarkably, the modulation depth of the exciton oscillator strength provided by these CAPs is among the largest ever experimentally observed. Comparable signals have only been reported in semiconductor nanostructures \cite{sun2000coherent, devos2007strong}, where the electron-phonon coupling is strongly enhanced by quantum confinement. We also notice that the detected exciton shift is one of the highest ever reached in condensed matter at RT under the influence of external perturbations (\textit{e.g.} pulsed light field, pulsed acoustic field, electric field, magnetic field), as summarized in Table 1. Comparable shifts were reported only for low-dimensional materials \cite{kuo2005strong}. Therefore, our observations point to the presence of a remarkably giant exciton-acoustic phonon coupling in RT bulk anatase TiO$_2$. An acousto-electric effect in this material is excluded as it is not piezoelectric. A surface polarization coming from an inhomogeneous distribution of oxygen vacancies at the surface of the material could provide an alternative explanation. However, this would imply that after excitation, the excitons would undergo a narrowing because of the screening of the built-in field \cite{cartwright1993magnitude}. This is not occurring in the present case, and the excitons are undergoing huge broadening due to long-range Coulomb screening. Therefore, this contribution, if at all present, is negligible compared to other generation mechanisms, which we now discuss. Indeed, hereafter, the details of the coupling are modeled by many-body perturbation theory, which nicely reproduces the experimental results. Such an advanced \textit{ab initio} treatment of the exciton-phonon coupling in bulk solids has never been explored before and it shows that the DP is the origin of the giant electron-phonon coupling in anatase TiO$_2$.

The description of transient reflectivity signals of acoustic origin is based on the perturbative approach developed by Thomsen et al. \cite{ref:thomsen_prb}. Here, the change in the reflectivity produced by the acoustic strain reads
\begin{eqnarray}
\label{drr}
\frac{\delta R}{R} = 2\operatorname{Re}\Big\{{ \frac{4ik\tilde{n}}{1-\tilde{n}^{2}}\frac{d\tilde{n}}{d\eta}\int_{0}^{\infty}\eta(z,t)e^{2ik\tilde{n}z}dz}\Big\},
\end{eqnarray}
where $\operatorname{Re}$ denotes the real part, $z$ = 0 defines the crystal surface, $k$ is the probe light wave vector in vacuum and $\tilde{n} = n_1 + in_2$ is the complex refractive index. This expression is governed by the time-dependent spatial overlap integral between the longitudinal CAP strain field, $\eta(z,t)$, and the electric field of the backscattered probe light. The exciton-phonon coupling strength is embodied by the photoelastic coefficient $d\tilde{n}/d\eta$, which is linked to the DP matrix element. Importantly for our discussion, in the original Thomsen model the photoelastic coefficient is assumed to be independent of the strain, thus leading to an opposite but equivalent shift of the material optical spectrum under the application of compressive or tensile stress.

\begin{figure}[t]
	\begin{center}
		\includegraphics[width=0.8\columnwidth]{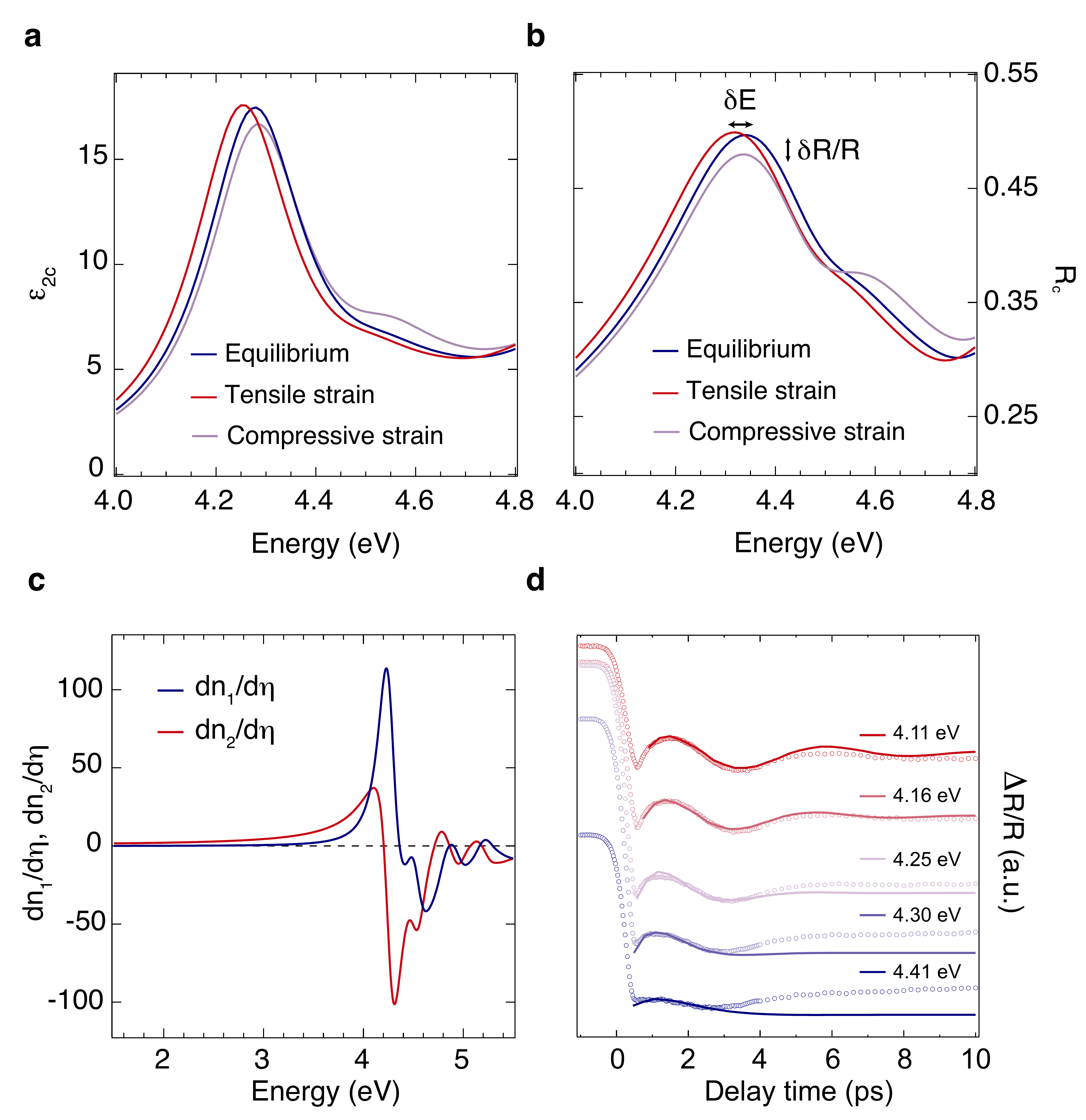}
		\caption{(a) Calculated imaginary part of the dielectric function in the BSE-GW scheme for the equilibrium unit cell (blue curve) and in the presence of a 0.2\% tensile (red curve) and compressive (violet curve) strain (b) Calculated reflectivity in the BSE-GW scheme for the equilibrium unit cell (blue curve) and in the presence of tensile (red curve) and compressive (violet curve) strain. (c) Calculated real (blue curve) and imaginary (red curve) parts of the photoelastic coefficient. (d) Simulated transient acoustic response at different probe photon energies (solid lines), as indicated in the labels. The dotted lines are the experimental data.}
		\label{fig:Fig3}
	\end{center}
\end{figure}

To evaluate Eq. (1), the photoelastic coefficients are typically computed from band structure calculations. However, this becomes challenging when the material optical properties are governed by strong excitonic correlations. In Refs. \cite{ref:baldini_TiO2, baldini_rutile}, we showed that many-body perturbation theory provides a very accurate description of the electronic and optical properties of TiO$_2$ single crystals. Here, we extend this approach to the case of strained anatase TiO$_2$, and extract the fundamental electron-phonon matrix elements that are relevant for both the generation (photoinduced strain $\eta(z,t)$) and the detection (photoelastic coefficient $d\tilde{n}/d\eta$) mechanisms. Specifically, the single-particle excitation spectrum of the material is calculated at the GW level while applying a 0.2\% deformation of the unit cell along the [010] axis. This deformation is artificially introduced to mimic the photoinduced strain propagating in the crystal under our experimental conditions, and to study its effects on the single-particle electronic structure. In a second step, we compute the optical spectrum in the presence of many-body electron-hole correlations by solving the BSE, and study how the excitons are renormalized by the macroscopic strain field. We remark that our theoretical analysis is performed in the case of a pristine TiO$_2$ crystal, taking into account only direct optical transitions and neglecting the additional screening induced by free carriers and the presence of indirect (phonon- or impurity-assisted) transitions. Although the latter effects are not expected to modify our conclusions (as discussed in Ref. \cite{ref:baldini_TiO2}), they could be responsible for some of the remaining discrepancies between theory and experiment. The GW approach allows us to refine the values of the DPs obtained by standard density-functional theory (DFT) \cite{ref:yin_effective}. Our estimate of the DPs experienced by the lowest states in the valence ($d_h$) and conduction band ($d_e$) yields $d_h$ = -0.066 eV/GPa and $d_e$ = -0.096 eV/GPa. The negative sign of the DPs suggests that the photoinduced stress is compressive in nature. 
This allows us to evaluate the efficiency of the DP mechanism in comparison to the other dominant generation process in non-magnetic and non-piezoelectric materials, which is the thermoelasic mechanism (phonon pressure). From our analysis (see the Supplementary Information, SI), we conclude that the DP mechanism provides a contribution $\sim$20 times larger than that produced by thermoelasticity. This result can be obtained only with a reliable estimate of the DPs, as the one provided at the GW level. 

\begin{table}
	\begin{center}
		\footnotesize
		\begin{tabular}{cccccc}
			\hline
			Material & Dimensionality & Perturbation & Temperature & Exciton Shift & Ref.\\
			\hline
			
			Anatase TiO$_2$ & Bulk & Pulsed strain field	&	295 K & 30-50 meV &	This work\\

			Zn$_{1-x}$Cd$_{x}$Se & Quantum wells	& Pulsed light field	&	10 K & 4 meV &	\cite{gupta2001ultrafast}\\
			
			WS$_2$, WSe$_2$ & Monolayer	& Pulsed light field	&	295 K & 10-18 meV &	\cite{kim2014ultrafast, sie2015valley}\\
			
			CH$_3$NH$_3$PbI$_3$  & Bulk & Pulsed light field	&	295 K & 10 meV	&	\cite{yang2016large}\\
			
			GaAs/AlGaAs & Heterostructure	& Pulsed strain field	&	1.8 K & 1 meV &	\cite{akimov2006ultrafast}\\
			
			ZnSe/ZnMgSSe & Quantum wells & Pulsed strain field	&	1.8 K & 10 meV &	\cite{scherbakov2007chirping}\\
			
			CdSe & Quantum dots	& Pulsed strain field	&	295 K & $<$ 1 meV &	\cite{sagar2008size}\\
			
			ZnO, CdS, ZnSe & Bulk	& Strain field	&	1.8 K & 7-20 meV &	\cite{langer1970spin}\\
			
			GaAs/AlGaAs & Quantum wells	& Electric field	&	295 K & 22 meV &	\cite{miller1984band}\\
			
			Ge/SiGe & Quantum wells	& Electric field	&	295 K & 30 meV &	\cite{kuo2005strong}\\
			
			WS$_2$, MoS$_2$ & Monolayer	& Magnetic field	&	4 K & 8 meV &	\cite{stier2016exciton}\\
			
		\end{tabular}
		\caption{Exciton shifts in different materials under distinct external perturbations.}
		\label{tab:globalfit1}
	\end{center}
\end{table}

Thereafter, we compute the dielectric function and the reflectivity in the case of the unstrained unit cell and in presence of a tensile/compressive strain along the [010] axis (Fig. S2 and Fig. 3(a,b)). We find that the exciton peak strongly reacts to the strain field, undergoing a dramatic renormalization of its energy and a pronounced modulation of its intensity. Remarkably, the peak energy shift ($\delta$E = 30 meV) and intensity change ($\delta$R/R = 3.5 \%) in the reflectivity show excellent matching with the experimental data. Moreover, the observed exciton behavior suggests that the photoelastic coefficients $dn_1/d\eta$ and $dn_2/d\eta$ are not symmetric upon application of compressive or tensile strain (see Fig. S3). Specifically, in the presence of a tensile strain, the exciton peak undergoes a larger shift towards the red; in contrast, upon the application of a compressive strain, the exciton displays higher resistence to blueshift. This intriguing aspect is not accounted for by the original Thomsen model of ultrafast acoustics, where the photoelastic coefficient is assumed to be independent of the strain and excitonic effects are totally neglected. This suggests that, in the present experimental conditions, the system is driven into a nonlinear (non-perturbative) regime where the photoinduced strain is large and the damped sinusoidal acoustic signal predicted by the Thomsen model may be distorted. As such, our results bear important similarities with the nonlinear propagation of light-induced solitons \cite{scherbakov2007chirping} and shock-waves \cite{bojahr2012calibrated} in materials subject to giant strain values, and show that the interaction between excitons and phonons beyond the linear regime can give rise to an  anomalous renormalization of the excitonic states. To simplify the modelling of our data within the perturbative regime, we calculate the photoelastic coefficients using the central finite differences method (Fig. 3(c)). Large values are found in the vicinity of the exciton resonance, indicating that the electron-phonon coupling is effective only in this photon energy range.

Finally, we evaluate Eq. (1) using the parameters computed above. We consider that the photoinduced DP stress is produced within the skin depth of the pump field ($\xi=\lambda/4\pi n_2$ $\sim$ 12 nm). Since the crystal is semi-infinite, the resulting photoinduced strain pulse is assumed to be bipolar, with $\eta(z,t)=-\eta_0 \sign(z-vt)e^{-\mid{ z-vt}\mid / \xi}$ ($v$ is the longitudinal sound velocity). Based on the computed DPs and elastic properties (a bulk modulus $B$ = 181 GPa), the photoinduced strain along the [010] direction is $\eta_0 = (d_e+d_h)BN/\rho_0v^2 \sim 10^{-3}$ ($\rho_0$ is the mass density). The final calculation is performed with the computed photoelastic coefficient of Fig. 3(c). Our model yields damped sinusoidal functions that are shown in Fig. 3(d) and compared with the experimental traces. We observe that the oscillatory response is well reproduced and the calculated frequencies ($\mathrm{\nu_{{ph},th}}$, Fig. 2(d)) are in excellent agreement with the experimental ones. Small deviations in the oscillation amplitude (by a factor of 0.6$\div$1.2) and damping are caused by the distortion that the signal undergoes beyond the linear propagation regime. More importantly, some of the traces were multiplied by -1 to match the experimental curves, implying that our model does not reproduce the phase of the oscillation over the whole probed range. Indeed, the phase depends on the ratio between the real and imaginary part of the photoelastic coefficient, which are here calculated with the method of the central differences to mimic the perturbative regime of the Thomsen model. Thus, the disagreement between the phase observed in our experimental data and the one predicted by the Thomsen model shows how the latter describes the CAP propagation in a phenomenological and approximate way. Despite this limitation, the overall agreement between our data and the model is exceptional, and underlines how crucial our \textit{ab initio} approach is to obtain realistic CAP signals. We underline that the optical spectra calculated with the random phase approximation on top of the GW results do not provide realistic signals \cite{ref:baldini_TiO2}, leading to much smaller photoelastic coefficients than the experimental ones (see SI). This highlights the importance of solving the BSE to address the impact of the lattice motion on the excitonic states, thus opening new perspectives for the accurate prediction and design of tailored exciton-phonon coupling schemes. Future extensions of the Thomsen model to the non-perturbative regime and in the presence of strong exciton nonlinearities will reveal the complex structure of the electron-phonon coupling in strongly interacting materials such as anatase TiO$_2$.

In conclusion, we combined state-of-art ultrafast broadband deep-UV spectroscopy and many-body perturbation theory calculations to elucidate the details of the coupling between excitons and CAPs in the technologically relevant oxide anatase TiO$_2$. Efficient generation and detection of CAPs in the 100-200 GHz range is observed at RT in the vicinity of an exciton resonance, which is characterized by photoelastic coefficients as large as those found in the visible range for quantum confined nanostructures and an exceptional exciton band tuning of 30-50 meV with a moderate strain value. This result assumes a particular importance when compared to the coherent manipulation of excitons via the light field itself within the framework of the optical Stark effect, which so far has produced RT exciton shifts over the visible range of 10-20 meV in transition metal dichalcogenides \cite{kim2014ultrafast, sie2015valley} and 2 meV in hybrid perovskites \cite{yang2016large}. Thus, our findings open intriguing perpectives for deep-UV acousto-optics and acousto-excitonics, in which the additional electronic contribution to the exciton renormalization can be suppressed by engineering a source of CAPs (\textit{e.g.} a metallic film) on top of an anatase TiO$_2$ crystal. Finally, in analogy to the growing field of ``active plasmonics" \cite{macdonald2009ultrafast}, we envision the rise of ``active excitonics" schemes in which the excitonic transport can be selectively modulated at high-speeds by tailored opto-acoustic stimuli.

\section*{Methods}

\subsection{Single crystal growth and characterization}
\label{Methods_SingleCrystals}

High-quality single crystals of anatase TiO$_2$ were produced by a chemical transport method from anatase powder and NH$_4$Cl as transport agent, similar to the procedure described in Ref. \cite{berger1993growth}. In detail, 0.5 g of high-purity anatase powder were sealed in a 3 mm thick, 2 cm large and 20 cm long quartz ampoule together with 150 mg of NH$_4$Cl, previously dried at $60\,^{\circ}\mathrm{C}$ under dynamic vacuum for one night, and 400 mbar of electronic grade HCl. The ampoules were placed in a horizontal tubular two-zone furnace and heated very slowly to $740\,^{\circ}\mathrm{C}$ at the source, and $610\,^{\circ}\mathrm{C}$ at the deposition zone. After two weeks, millimeter-sized crystals with a bi-pyramidal shape were collected and cut into rectangular bars (typically 0.8 $\times$ 0.6 $\times$ 0.15 $\mathrm{mm^3}$). The doping level was determined via ARPES or transport measurements to be $n$ = 2 $\times$ 10$^{19}$ cm$^{-3}$.

\subsection{Ultrafast broadband deep-UV spectroscopy}
\label{Methods_Ultrafast}

The ultrafast optical experiments were performed using a novel set-up of tunable deep-ultraviolet (UV) pump and broadband UV probe, described in detail in Ref. \cite{aubock2012ultrabroadband}. A 20 kHz Ti:Sapphire regenerative amplifier (KMLabs, Halcyon + Wyvern500), providing pulses at 1.55 eV, with typically 0.6 mJ energy and around 50 fs duration, pumped a noncollinear optical parametric amplifier (NOPA) (TOPAS white - Light Conversion) to generate sub-90 fs visible pulses (1.77 - 2.30 eV range). The typical output energy per pulse was 13 $\upmu$J. Around 60\% of the output of the NOPA was used to generate the narrowband pump pulses. The visible beam, after passing through a chopper, operating at 10 kHz and phase-locked to the laser system, was focused onto a 2 mm thick BBO crystal for nonlinear frequency doubling. The pump photon energy was controlled by the rotation of the crystal around the ordinary axis and could be tuned in a spectral range up to $\sim$0.9 eV ($\sim$60 nm) wide. For our purpose, the pump photon energy was set at 4.50 eV, in order to selectively perturb the spectral region above the c-axis excitonic peak of anatase TiO$_2$. The typical pump bandwidth was 0.02 eV (1.5 nm) and the maximum excitation energy was about 120 nJ. The pump power was recorded on a shot-to-shot basis by a calibrated photodiode for each pump photon energy, allowing for the normalization of the data for the pump power. The remaining NOPA output was used to generate the broadband UV probe pulses with $\sim$1.3 eV ($\sim$100 nm) bandwidth through an achromatic doubling scheme.

For studying the anatase TiO$_2$ single crystals, the set-up was used in the reflection geometry. The specimens were mounted on a rotating sample holder, in order to explore the transient reflectivity ($\Delta$R/R) along the desired crystalline axis. Pump and probe pulses, which have the same polarization, were focused onto the sample, where they were spatially and temporally overlapped. The typical spot size of the pump and the probe were 100 $\upmu$m and 40 $\upmu$m full-width at half-maximum respectively, resulting in a homogeneous illumination of the probed region. The portion of the probe beam reflected by the surface of the crystal was detected and the time evolution of the difference in the UV probe reflection with and without the pump pulse reconstructed. After the sample, the reflected probe was focused in a multi-mode optical fiber (100 $\upmu$m), coupled to the entrance slit of a 0.25 m imaging spectrograph (Chromex 250is). The beam was dispersed by a 150 gr/mm holographic grating and imaged onto a multichannel detector consisting of a 512 pixel CMOS linear sensor (Hamamatsu S11105, 12.5 $\times$ 250 $\upmu$m pixel size) with up to 50 MHz pixel readout, so the maximum read-out rate per spectrum (almost 100 kHz) allowed us to perform shot-to-shot detection easily. For the reported measurements, the set-up was operated with a time resolution of 700 fs. All the experiments were performed at RT. 

\subsection{Many-body perturbartion theory calculations}
\label{AbInitio_Details}

Many-body perturbation theory at the level of the GW and the Bethe-Salpeter Equation (BSE) \cite{ref:hedin1} was employed to compute the band structure and the dielectric response of bulk anatase TiO$_2$. The GW and BSE calculations were performed on-top of eigenvalues and eigenfunctions obtained from DFT. We used the planewave pseudopotential implementation of DFT as provided by the package Quantum Espresso. GW and BSE calculations were performed with the BerkeleyGW package \cite{ref:deslippe}.

The DFT calculations were performed using the generalized gradient approximation (GGA) as in the Perdew-Burke-Ernzerhof (PBE) scheme for the exchange-correlation functional. The Ti norm-conserving pseudopotential was generated in the Rappe-Rabe-Kaxiras-Joannopoulos (RRKJ) scheme \cite{ref:RRKJ}, including semicore 3\textit{s} and 3\textit{p} states. While standard structural and electronic quantities are already converged in DFT  with an energy cutoff of 90 Ry, the energy cutoff used here was raised to 160 Ry to properly include the high number of bands necessary to reach convergence for the many-body evaluated properties. Bulk anatase TiO$_2$ was modeled on a body-centered tetragonal lattice containing 2 Ti atoms and 4 O atoms (primitive cell) with lattice parameters (optimized at the PBE level) a = b = 3.79 \AA~ and c = 9.66 \AA. The experimental lattice constants at RT are a = b = 3.78 \AA~ and c = 9.51 \AA. Scaling these parameters to zero temperature via a linear extrapolation of the temperature dependence of the lattice constant at high temperature, appearing in Ref. \cite{rao1970thermal}, yields a = b = 3.78 \AA~ and c = 9.49 \AA.   

The ground state electronic density is properly described with a coarse 4$\times$4$\times$4 $k$-point grid for sampling of the BZ. The GW quasiparticle corrections to the DFT eigenvalues were performed at the one-shot level of theory (G$_0$W$_0$). For the computation of the polarizability and inverse dielectric matrices in BerkeleyGW, we employed a total of 2474 CBs and G-vectors with kinetic energies up to 46 Ry, whereas the self-energy operator was computed using 2472 unoccupied bands and a G-vector cutoff energy of 46 Ry and 160 Ry for the screened and bare Coulomb matrices, respectively. The coarse 4$\times$4$\times$4 $k$-point grid sampling is sufficient for the description of the quasiparticle corrections, while a high number of bands is mandatory to get a proper description of screening effects and many-body corrections. The electronic band structure was finally obtained by interpolating GW corrections on top of a more refined DFT calculation with a 16$\times$16$\times$16 grid. The fully converged BSE results shown in the main text were obtained with BerkeleyGW. We used a shifted grid with up to 16$\times$16$\times$16 $k$-points (4096 irreducible $k$-points). The six lowest CBs and six topmost VBs were included to solve the excitonic Hamiltonian. Spin-polarized calculations were performed to highlight possible dark excitons due to triplet excitations but no measurable differences with respect to the spin-restricted results were obtained. More details are provided in Ref. \cite{ref:baldini_TiO2}.

To estimate the role of the electron-acoustic phonon coupling in the electronic and optical properties of anatase TiO$_2$, we performed frozen phonon DFT + GW + BSE calculations by applying a strain of 0.2\% along the [010] crystallographic direction. The results of these calculations are shown in Fig. 3(a-c) of the main text.

\subsection{Data availability}

The data that support the findings of this study are available from the corresponding authors upon reasonable request.

\subsection{Acknowledgements}
We thank Marco Grioni and Simon Moser for providing the sample used for this study, Letizia Chiodo, Maurizia Palummo and Christian Bernhard for valuable discussions. We acknowledge support by the Swiss NSF via the NCCR:MUST and R'EQUIP, and by the European Research Council Advanced Grant DYNAMOX. This project has received funding from the European Union’s Horizon 2020 research and innovation programme under the Marie Sklodowska-Curie grant agreement No 753874.

\subsection{Authors contributions}
E.B. performed the experiments, A.D. and A.R. performed the theoretical calculations, E.B., T.P., and O.C. analyzed the experimental data, E.B. and P.R. contributed to the data interpretation, E.B., P.R. and M.C. wrote the final manuscript. All authors participated in the final version of the article. E.B. and M.C. conceived the study.

\subsection{Author information}
The authors declare having no competing financial interests. Correspondence and requests for materials should be addressed to E.B. (email: ebaldini@mit.edu) or M.C. (email: majed.chergui@epfl.ch).

\clearpage
\newpage

\setcounter{section}{0}
\setcounter{figure}{0}
\renewcommand{\thesection}{S\arabic{section}}  
\renewcommand{\thetable}{S\arabic{table}}  
\renewcommand{\thefigure}{S\arabic{figure}} 
\renewcommand\Im{\operatorname{\mathfrak{Im}}}
\titleformat{\section}[block]{\bfseries}{\thesection.}{1em}{} 

\section{S1. Global fit analysis}

To describe the significant parameters of the $\Delta$R/R response, we performed a global fit analysis, by selecting thirteen temporal traces at different photon energies. Since our experimental data were measured up to 1 ns, our global fit analysis covered this long temporal window. The temporal traces comprise two main signals: i) The incoherent response caused by electron-hole generation and relaxation; and ii) The strongly damped sinusoidal oscillation caused by the propagation of coherent acoustic phonons (CAPs). These two contributions display very different weights across the probed spectral range: The incoherent signal is peaked around the c-axis exciton feature at 4.26 eV, whereas the coherent signal shows its maximum amplitude around 4.11 eV and decreases its weight with increasing probe photon energy. A satisfactory fit could be obtained by using four exponential functions (with relaxation time $\tau_i$) and a damped sinusoidal term (with frequency $\Omega$, damping $\tau_D$ and phase $\phi$) convolved with a Gaussian response $I(t)$ accounting for the temporal shape of the pump pulse
\begin{equation}
f(t) = I(t) * \sum_{i=1}^4 A_i e^{-t/\tau_i} + B \sin(\Omega t + \phi) e^{-t/\tau_{D}}.
\label{Fit_TiO2_CAP}
\end{equation}
The parameters extracted from the global fit analysis are listed in Table S2.\\

	\begin{table}[h!]
	\begin{center}
		\normalsize
		\begin{tabular}{ccc}
			\hline
			Parameter & Value\\
			\hline
			$\tau_1$              & (7.4 $\pm$ 0.2) ps\\
			$\tau_2$              & (35 $\pm$ 2) ps\\
			$\tau_3$              & (240 $\pm$ 20) ps\\
			$\tau_4$             & (3.7 $\pm$ 0.5) ns\\
			$\tau_D$              & (2.28 $\pm$ 0.02) ps\\
			$\omega$              & (1.82 $\pm$ 0.01) rad/sec\\
			\hline
		\end{tabular}
		\caption{Parameters extracted from the first global fit.}
		\label{tab:globalfit1}
	\end{center}
\end{table}

\section{S2. Generation mechanism}

To quantify the electronic contribution to the photoinduced stress ($\sigma_{DP}$), we rely on the following expression
\begin{equation}
\sigma_{DP} = \sum_k \delta N(k) \frac{dE_k}{d\eta},
\label{sigma_DP}
\end{equation}
\noindent where $\delta N$($k$) is the change of the electronic concentration at level \textit{k} and d$E_{k}$/d$\eta$ is the deformation potential parameter. Given that after 50 fs (\textit{i.e.} a much faster timescale than the detected CAP period) the carriers have thermalized to the bottom of the respective bands at $\Gamma$ and $\sim$ X \cite{baldini2017exciton}, the expression can be simplified as
\begin{equation}
\sigma_{DP} = -N B \Bigg[\frac{dE_e}{dP}\bigg|_{\Gamma} + \frac{dE_h}{dP}\bigg|_{X}\Bigg] = - N B (d_e + d_h),
\label{sigma_DP_2}
\end{equation}
\noindent where $N$ is the photoinduced carrier concentration and $B$ is the bulk modulus. 

However, due to the extremely fast intraband relaxation of less than 50 fs \cite{baldini2017exciton}, the electronic pressure can compete with the phononic pressure for the CAP excitation process. The phononic contribution to the photoinduced stress ($\sigma_{TE}$) can be written as
\begin{equation}
\sigma_{TE} = -\alpha_V B \Delta T_L = -\alpha_V B N E_{exc}/C_L,
\label{sigma_TE}
\end{equation}
\noindent where $\alpha_V$ is the volumetric thermal expansion coefficient. For a tetragonal crystal, $\alpha_V = 2\alpha_{\perp} + \alpha_{\parallel}$, where $\alpha_{\perp}$ and $\alpha_{\parallel}$ are the in-plane and  the out-of-plane thermal expansion coefficients, respectively. $\Delta T_L$ is the lattice temperature, $C_L$ is the lattice heat capacity per unit volume and $E_{exc}$ is the excess energy with respect to the optical bandgap energy. The ratio $\sigma_{DP}$/$\sigma_{TE}$ reads
\begin{equation}
\frac{\sigma_{DP}}{\sigma_{TE}} = \frac{C_L (d_e + d_h)}{\alpha_V  E_{exc}}
\label{sigma_DP_TE}
\end{equation}

Substituting the computed values of $d_h$ and $d_e$ yields $\sigma_{DP}$/$\sigma_{TE}$ = -27.47. Thus, we conclude that the deformation potential mechanism provides the dominant contribution to the generation of the observed CAPs. Importantly, this result can be obtained only with a reliable estimate of the DPs, as the one provided at the GW level. In contrast, when relying on the generalized-gradient approximation level of density-functional theory \cite{ref:yin_effective}, the value $\sigma_{DP}$/$\sigma_{TE}$ = -4.2 is found.

\section{S3. Perturbative model for coherent acoustic phonons}

To simulate the change in the sample reflectivity produced by the propagating acoustic strain ($\delta$R/R), we rely on the perturbative approach developed in Ref. \cite{ref:thomsen_prb}. Specifically, we make use of an expression that describes the time-dependent spatial overlap integral of the longitudinal CAP strain field along the [010] direction ($\eta(z,t)$) with the back scattered light probe electric field
\begin{eqnarray}
\label{drr}
\frac{\delta R}{R} = 2\operatorname{Re}\Big\{{ \frac{4ik\tilde{n}}{1-\tilde{n}^{2}}\frac{d\tilde{n}}{d\eta}\int_{0}^{\infty}\eta(z,t)e^{2ik\tilde{n}z}dz}\Big\},
\end{eqnarray}
where $\operatorname{Re}$ denotes the real part, $z$ = 0 defines the TiO$_2$ surface, $k$ is the probe light wave vector in vacuum and $\tilde{n} = n_1 + in_2$ is the complex refractive index. In this expression, the exciton-phonon coupling parameter is represented by the photoelastic coefficient $d\tilde{n}/d\eta$, which can be written as
\begin{equation}
\frac{d\tilde{n}}{d\eta} = \frac{dn_1}{d\eta} + i\frac{dn_2}{d\eta} = \frac{dE}{d\eta}\Bigg(\frac{dn_1}{dE} + i\frac{dn_2}{dE}\Bigg).
\end{equation}
Here, the quantity $dE/d\eta$ represents the deformation potential parameter and is assumed to be independent of the probe photon energy and the strain itself. To compute the above expression over the probed spectral range, the pulse strain profile is considered to have a bipolar shape, as if carrier diffusion is negligible during the timescale of the detected acoustic field. The sound velocity $v$ = 9100 m/s is taken from the literature \cite{shojaee2009first}, while the other quantities are calculated from first principles using our many-body perturbation theory approach. The results are shown in Fig. 3(d) of the main text, in which we compare the experimental $\Delta$R/R traces with the simulated signal from the CAP. Moreover, we multiplied by -1 some of the traces to match the experimental curves. This implies that our calculation does not reproduce the phase of the signal in a specific portion of the spectrum, which is due to an only partial agreement between the theoretical photoelastic coefficient and the experimental one. Figure S1 shows the computed acoustic response in an enlarged scale (dotted lines), together with its fit to a damped sinusoidal function (solid lines). The frequency extracted from the fit to the computed traces are shown in Fig. 2(d) of the main text. 

\begin{figure}[h!]
	\begin{center}
		\includegraphics[width=0.6\columnwidth]{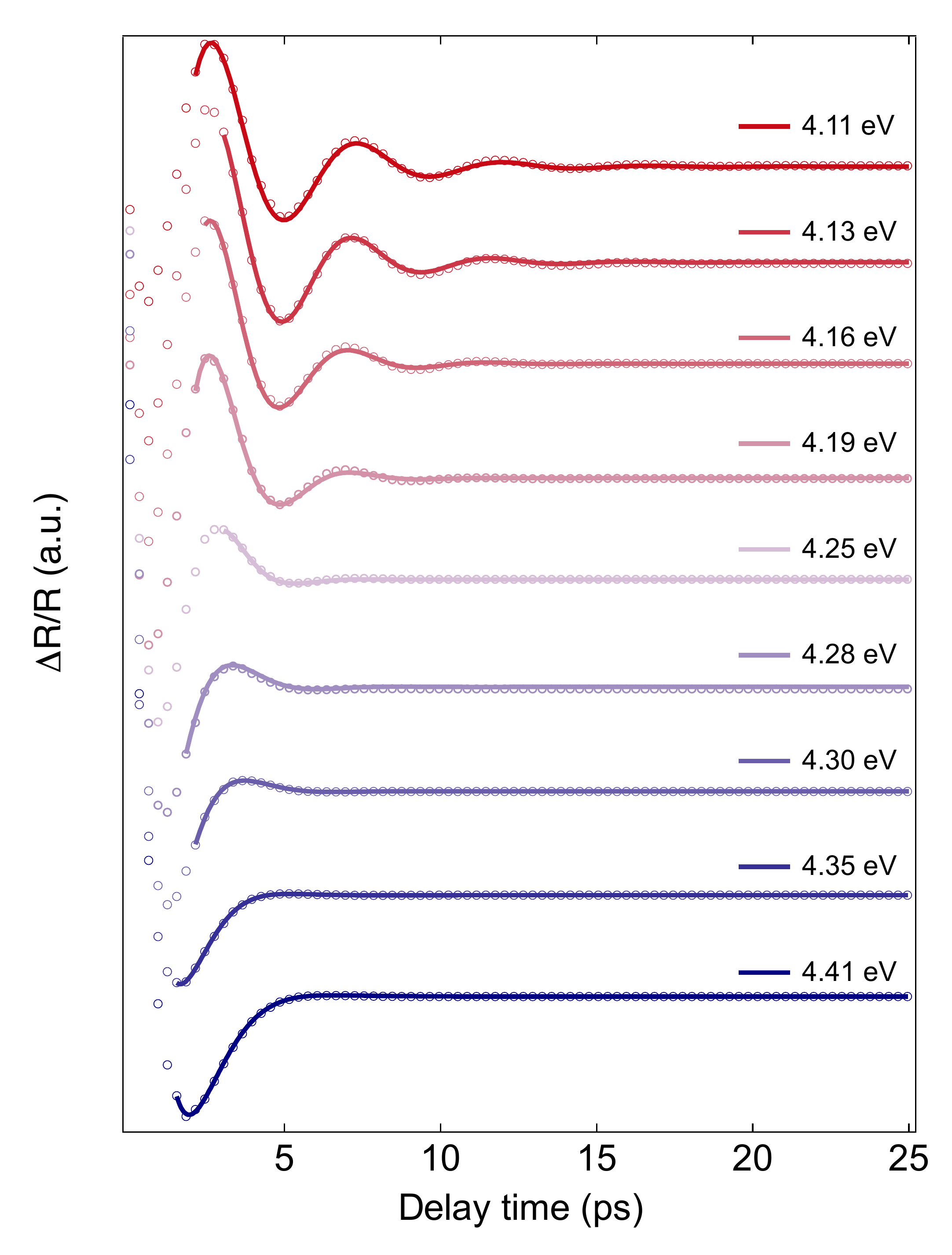}
		\caption{Computed acoustic response (dotted lines), together with its fit to a damped sinusoidal function (solid lines). The probe photon energies are indicated in the labels.}
		\label{fig:FigS1}
	\end{center}
\end{figure}
\newpage

\begin{figure}[h!]
	\begin{center}
		\includegraphics[width=\columnwidth]{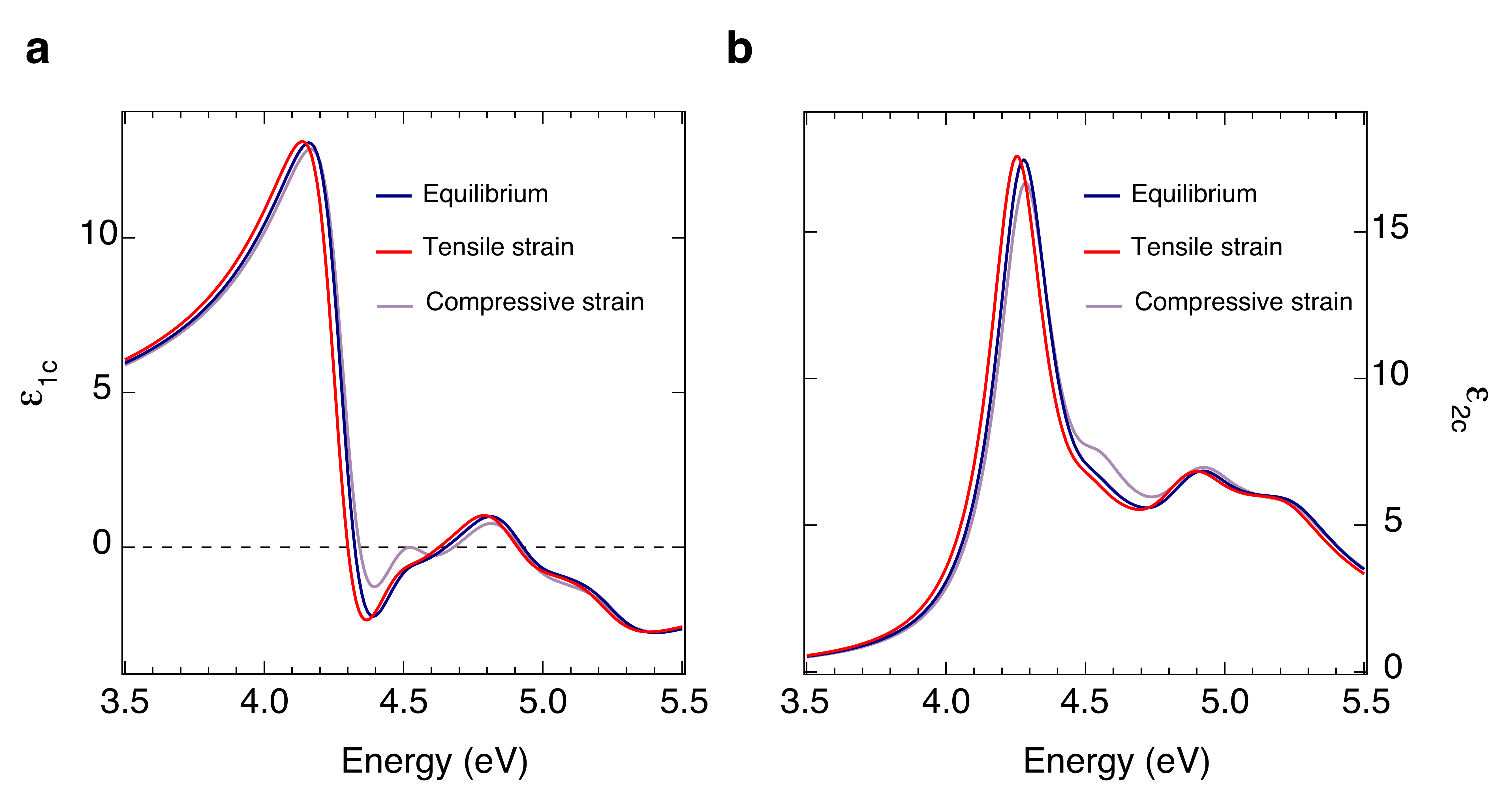}
		\caption{Calculated (a) real and (b) imaginary part of the dielectric function in the BSE-GW scheme for the equilibrium unit cell (blue curve) and in the presence of a 0.2\% tensile (red curve) and compressive (violet curve) strain.}
		\label{fig:FigS2}
	\end{center}
\end{figure}

\begin{figure}[h]
	\begin{center}
		\includegraphics[width=\columnwidth]{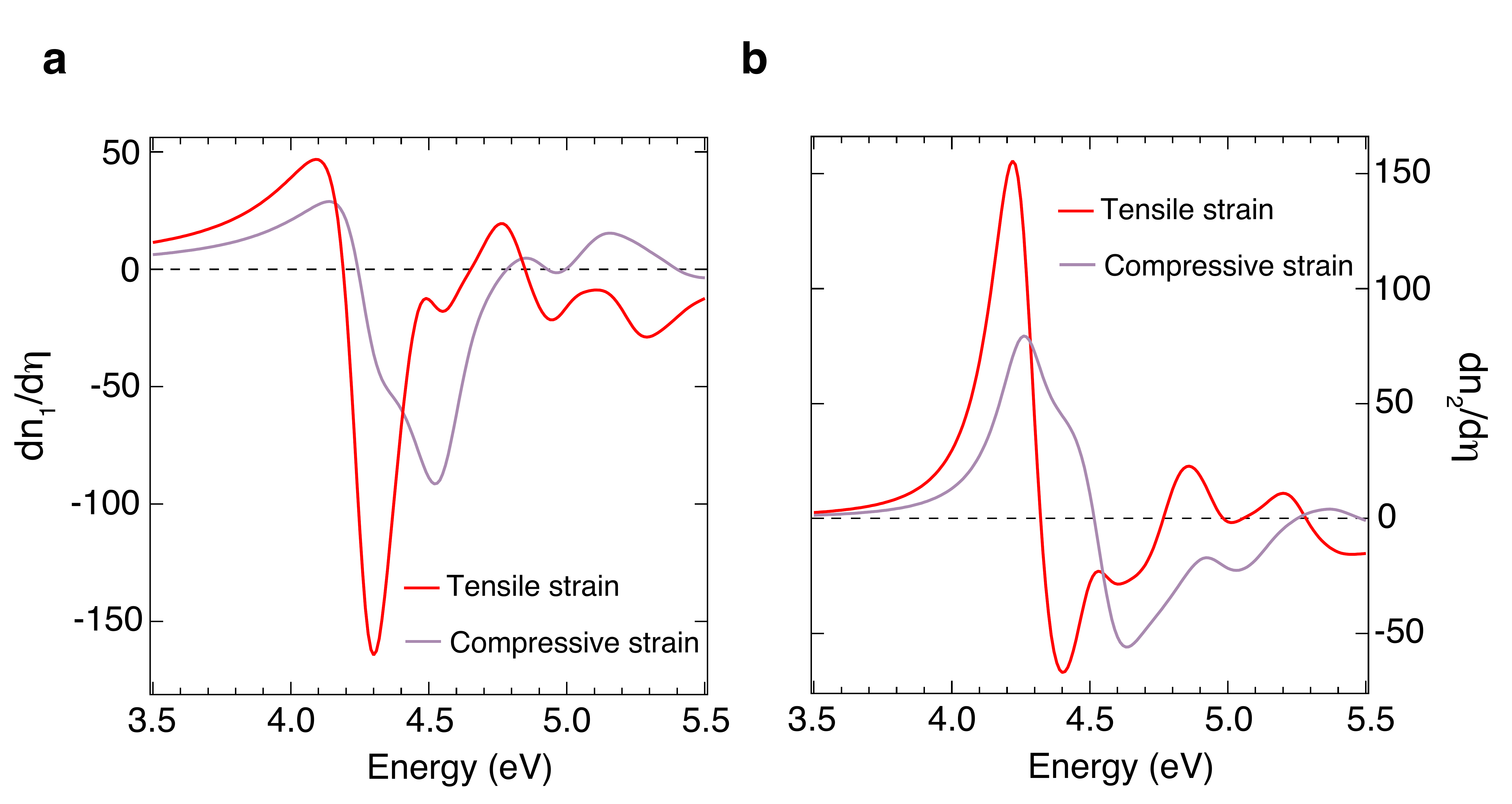}
		\caption{Calculated (a) real and (b) imaginary part of the photoelastic coefficient in the presence of a 0.2\% tensile (red curves) and compressive (violet curves) strain. The red (violet) curves have been obtained using the method of the backward (forward) differences.}
		\label{fig:FigS3}
	\end{center}
\end{figure}

\begin{figure}[h]
	\begin{center}
		\includegraphics[width=0.8\columnwidth]{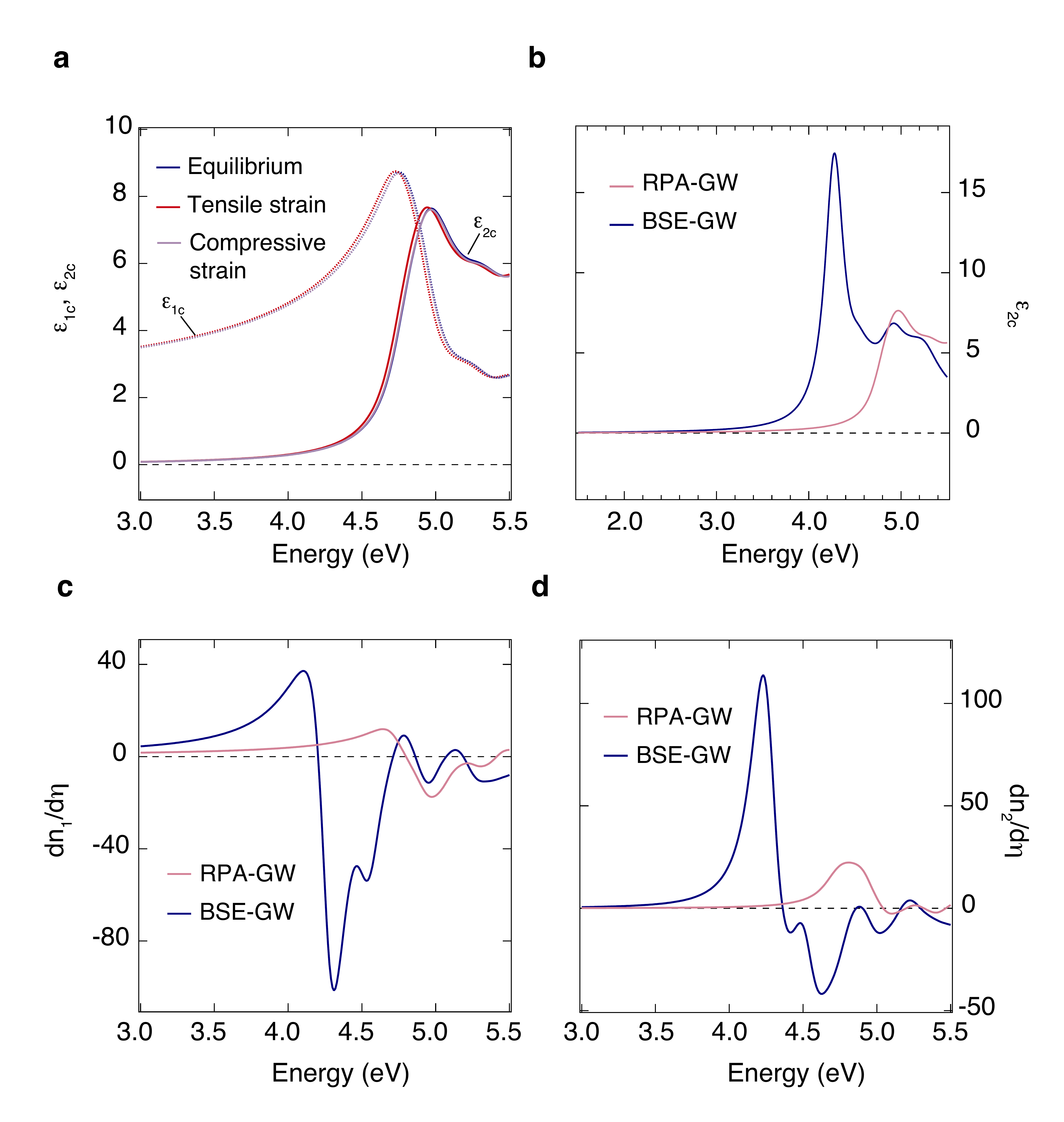}
		\caption{(a) Calculated real (dashed lines) imaginary (solid lines) parts of the dielectric function in the RPA-GW scheme for the equilibrium (blue curve) and strained (red curve) unit cell. (b) Comparison between the imaginary part of the dielectric function calculated in the RPA-GW (blue curve) and BSE-GW (violet curve) schemes for the unstrained unit cell. (c) Calculated real part of the photoelastic coefficient in the RPA-GW (blue curve) and BSE-GW (violet curve) schemes. (d) Calculated imaginary part of the photoelastic coefficient in the RPA-GW (blue curve) and BSE-GW (violet curve) schemes. }
		\label{fig:FigS4}
	\end{center}
\end{figure}

\section{S4. Additional many-body perturbation theory calculations}

We solved the Bethe-Salpeter equation (BSE) on top of GW electronic structure calculations to compute the electrodynamic properties of pristine anatase TiO$_2$. Figure S2(a,b) shows the real and imaginary part of the dielectric function calculated in the case of the unstrained unit cell (blue curves), and in presence of a 0.2\% tensile (red curves) and compressive (violet curves) strain along the [010] axis, respectively. The calculation was perfomed over the 1.50-5.50 eV energy range, but we display the data between 3.50 eV and 5.50 eV for clarity. The resulting photoelastic coefficients $dn_1/d\eta$ and $dn_2/d\eta$ for the cases of tensile (red curves) and compressive (violet curves) strain are presented in Fig. S3(a,b), as computed with the method of backward and forward differences, respectively. We observe an asymmetry between the photoelastic coefficients, which suggests a departure from the assumption that the photoelastic coefficients are independent of the applied strain direction. 

We also verified that the solution of the Bethe-Salpeter equation (BSE) on top of GW electronic structure calculations is necessary to reproduce our experimental results. To this aim, we computed the optical properties of an equilibrium (unstrained) and of a strained anatase TiO$_2$ unit cell in the absence of excitonic correlations, \textit{i.e.} at the random-phase approximation (RPA) level on top of the same GW electronic structure calculations. Figure S4(a) shows the $\epsilon_{1c}$, $\epsilon_{2c}$ in the case of the unstrained (blue curve) and in presence of the strain along the [010] axis (red curve). Thereafter, we compared these results with those obtained at the BSE level of theory. Figure S4(b) shows that $\epsilon_{2c}$ in the RPA-GW scheme features a large energy gap, which does not find agreement with the experimental findings. These results are in accordance to the calculations reported in our previous study \cite{ref:baldini_TiO2}. Finally, in Fig. S4(c,d), we compare the photoelastic coefficients $dn_1/d\eta$ and $dn_2/d\eta$ within the two levels of theory, and observe that the shape and magnitude of the photoelastic coefficients undergo a strong renormalization when excitonic effects are taken into account. Importantly, our experimental results can be reproduced just within the BSE-GW level of theory.

	\newpage

\providecommand{\noopsort}[1]{}\providecommand{\singleletter}[1]{#1}%

\end{document}